\documentclass[article,twocolumn,amsmath,floats,showpacs,nofootinbib]{revtex4}

\usepackage{graphicx}% Include figure files

\usepackage{textcomp}
\usepackage{hyperref}

\hypersetup{colorlinks=true}

\begin{document}

\title{Spatially inhomogeneous discrete states of a superconductor upon injection of nonequilibrium quasiparticles from a point contact with a normal metal}

\author{I. K. Yanson, L. F. Rybal'chenko, V. V. Fisun, N. L. Bobrov, M. A. Obolenskii,* M. B. Kosmyna,** and V. P. Seminozhenko**}
\affiliation{B.I.~Verkin Institute for Low Temperature Physics and
Engineering, of the National Academy of Sciences
of Ukraine, prospekt Lenina, 47, Kharkov 61103, Ukraine
\\Email address: bobrov@ilt.kharkov.ua\\*A.M. Gorky State University. Kharkov;\\**Scientific-Industrial Complex "Monokristallreaktiv," Kharkov}
\published {(\href{http://fntr.ilt.kharkov.ua/fnt/pdf/14/14-11/f14-1157r.pdf}{Fiz. Nizk. Temp.}, \textbf{14}, 1157 (1988)); (Sov. J. Low Temp. Phys., \textbf{14}, 639 (1988)}
\date{\today}

\begin{abstract}The current-voltage characteristics (IVC) of $S-c-N$ point contacts of superconductors with a small coherence length ${{\xi }_{0}}$  reveal steps with discrete values of the differential resistance. This peculiarity is associated with a transition of the contact region of the superconductor to a spatially inhomogeneous state under the influence of the current injection of nonequilibrium quasiparticles penetrating the superconductor to a depth ${{l}_{E}}$ . The role of the relaxation of the disbalance between the occupancies of electron- and hole-like branches of the quasiparticle spectrum is manifested in the displacement of the position of the singularities on the IVC towards higher energies upon an increase in the magnetic field and/or temperature. This effect was observed in superconductors with different ratios of the contact diameter $d$  and ${{\xi }_{0}}$ or ${{l}_{E}}$ in the series $Ta\ (d\ll {{\xi }_{0}},\ {{l}_{E}})\to NbS{{e}_{2}}$, $N{{b}_{3}}Sn\ (d\gtrsim {{\xi }_{0}},\ {{l}_{E}})\to YB{{a}_{1.25}}S{{r}_{0.75}}C{{u}_{3}}{{O}_{7-\delta }}(d\gg {{\xi }_{0}},\ {{l}_{E}})$. Apparently, a jumplike displacement of the boundary between regions with suppressed and equilibrium values of the energy gap near the contact is responsible for the oscillations observed on the IVC of point contacts between single crystals of high- temperature superconductors (HTS) and a normal metal. The resistance-periodic step-like structure of the IVC allows us to estimate the penetration depth ${{l}_{E}}$ of the electric field in $NbS{{e}_{2}}$  and $YBaSrCuO$.

\end{abstract}

\maketitle

\section{INTRODUCTION}
The state of a superconductor in the contact region of $S-c-N$ or $S-c-S$ structures ($c$ stands for a constriction or microbridge) in the current regime has been attracting the attention of researchers for a long time. The simplest model \cite{Iwanyshyn} taking into account the violation of superconducting state by a critical current density or Joule heating assumes a local equilibrium between quasiparticle excitations, the condensate, and phonons. A similar model was worked out by the authors of Refs. \cite{Hahn1,Hahn2,Hahn3} for explaining the $dV/dI\left( V \right)$  oscillations of the $Ta-Ag$  point contacts. However, the investigations of metals by the point-contact spectroscopy methods showed that in most cases there is no local equilibrium between electrons and phonons in the current concentration region \cite{Yanson1}. Moreover, there is no equilibrium in a superconductor between quasiparticle excitations and the condensate, which is manifested in the disbalance between the population densities of the electron- and hole-like branches in the spectrum of quasiparticle excitations. It has been shown by Yanson et al. \cite{Yanson2,Yanson3} that the jumplike decrease in the excess current with increasing bias voltage $V$  for $S-c-N$ point contacts $Ta-Ta$, $Ta-Cu$, and $Ta-Au$ in which the spectral regime of current passage is observed in the entire interval of applied bias voltages, i.e., the inequality $d<{{l}_{\varepsilon }}$ , ${{\left( {{l}_{i}}\ {{l}_{\varepsilon }} \right)}^{{1}/{2}\;}}$  ($d$  is the contact diameter and ${{l}_{\varepsilon }}$ , ${{l}_{i}}$  are the inelastic and elastic electron mean free paths) associated with the attainment of the critical concentration by the nonequilibrium quasiparticles is observed and is essentially of nonthermal type. In highly pure $Ta-Cu$ contacts of size $d\ll {{\xi }_{0}}$  (${{\xi }_{0}}$  is the coherence length of the superconductor), this decrease is of the order of several percent and is accompanied by an increase in the differential resistance of the same order of magnitude.

In the present work, a steplike structure of the IVC is observed in $S-c-N$ type point contacts between the single crystals of $NbS{{e}_{2}}$, $YB{{a}_{1.25}}S{{r}_{0.75}}C{{u}_{3}}{{O}_{7-\delta }}$ and noble metals $Cu$ , $Ag$ . This structure is reminiscent of the IVC structure associated with the formation of the phase slip centers in thin superconducting filaments. Such a structure appears when ${{l}_{E}}$ is of the order of, or smaller than, $d$ . It is shown that the formation of this structure has a nonthermal origin and is associated with an essentially nonequilibrium state of the superconductor in the contact region. In the same way as the emergence of a nonequilibrium singularity on the IVC of tantalum point-contacts upon a decrease in temperature is associated usually with the characteristic energies $h{{\omega }_{ph}}$  of the phonon spectrum \cite{Yanson3}, the jumps in the excess current and differential resistance on the IVC appearing upon a temperature decrease of superconducting point contacts with a small ${{\xi }_{0}}$  are also fixed near certain characteristic energies $e{{V}_{ph}}$  corresponding to the characteristic energies of the phonon spectrum \cite{Bobrov} or other excitations with which the conduction electrons effectively interact \cite{Yanson4}. On the other hand, the multiplicity of the jumps in the differential resistance may lead to oscillating $dV/dI\left( V \right)$  dependences which are frequently observed in experiments \cite{Kirtley}.
\subsection{DISCUSSION OF EXPERIMENTAL RESULTS}
\begin{figure}[]
\includegraphics[width=8cm,angle=0]{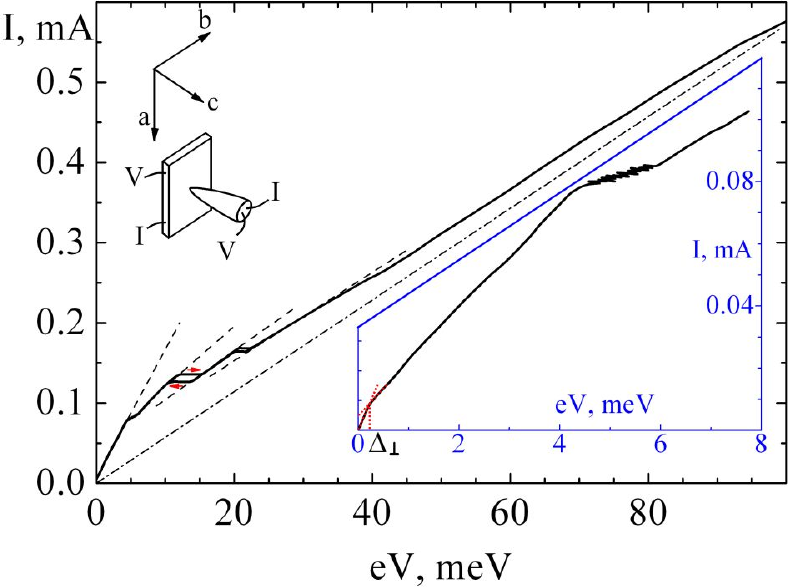}
\caption[]{Current-voltage characteristics of the $NbSe_2-Cu$ junction having a resistance $170\ \Omega $  at $T=1.68\ K$ . The dot-and-dash line shows the predicted IVC in the normal state, the dashed line corresponds to discrete values of differential resistance at the steps 60, 117, 150, and $170\ \Omega $. The upper inset shows the geometry of the experiment, and the lower inset shows the initial segement of the IVC (the dashed lines are the tangents to various segments of the IVC.}
\label{Fig1}
\end{figure}

\textbf{1.} $\textit{\textbf{NbS}}{{\textit{\textbf{e}}}_{\textit{\textbf{2}}}}$. Figure \ref{Fig1} shows the IVC of an $S-c-N$ point contact between a freshly cleaved single crystal of $NbS{{e}_{2}}$ and a copper needle oriented at right angles to the basal plane (see the upper inset to Fig.\ref{Fig1}).
\begin{figure}[]
\includegraphics[width=8cm,angle=0]{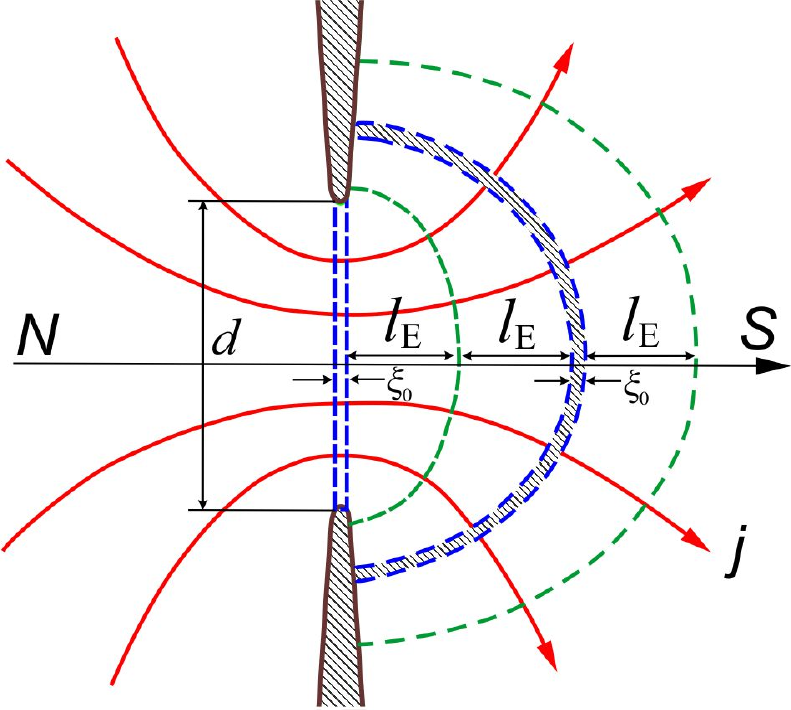}
\caption[]{The $S-c-N$ model of a point contact in the form of a circular hole of diameter $d$ in an impenetrable partition ($d>{{l}_{E}}>{{\xi }_{0}}$). The dashed curves are the boundaries between different nonequilibrium state regions of the superconductor, which are separated from the contact center by integral multiples of ${{l}_{E}}$.}
\label{Fig2}
\end{figure}
 For $eV>{{\Delta }_{\bot }}$ (${{\Delta }_{\bot }}\lesssim 0.4-0.6\ meV$   is the energy gap along the $c$-axis), the electric field penetrates the superconductor to a depth ${{l}_{E}}\approx {{l}_{\varepsilon }}$  (${{l}_{\varepsilon }}$  is the mean inelastic relaxation time for quasiparticles with energies $0<\varepsilon <eV$), and the differential resistance is apparently determined by the resistance of the ellipsoid with $z\approx {{l}_{E}}\lesssim d$ (Fig.\ref{Fig2}). This resistance is much higher than the zero-bias resistance (${{R}_{0}}=30\ \Omega $) (see the lower inset to Fig.\ref{Fig1}) associated with the normal face. In the model considered by us, it is significant that the contact diameter is small in comparison with the inelastic relaxation length of quasiparticles in the normal faces ($N$) for all bias voltages $eV$. For noble metals, this is obviously valid for $d\lesssim {{10}^{-5\ }}cm$. Hence, electrons (holes) are incident from the normal metal on the $N-S$  interface with an excess energy relative to the electrochemical pair potential in the superconductor (right up to $eV$).

Subsequent regions of IVC in Fig.\ref{Fig1} are reminiscent of peculiarities caused due to the formation of phase slip centers in a thin superconducting filament. The differential resistance ${{R}_{D}}$  changes abruptly on quasilinear segments of IVC: ${{R}_{1}}=62$, ${{R}_{2}}=117$, ${{R}_{3}}=150$, ${{R}_{4}}=174\ \Omega $. A consistent decrease in the value of the increment $\Delta {{R}_{D}}$ in successive steps is in qualitative agreement with the model with a jumpwise displacement of the ${S}'S$  boundary into the bulk of the superconductor (see Fig.\ref{Fig2}). Here, ${S}'$ is the superconducting region adjoining the constriction with a large nonequilibrium concentration of quasiparticle excitations into which the electric field penetrates, and $S$  is an equilibrium superconductor- corresponding to the state away from the constriction. In this model, we assume that the penetration depth ${{\lambda }_{L}}$  magnetic field into the superconductor is larger than $d$. It is obvious that even for voltages that exceed the region where the steps are observed, the excess current on the IVC continues to be finite and nearly constant. This means that the order parameter (energy gap) in the contact region ${S}'$ also has a finite value whose order of magnitude is equal to its unperturbed value.
\begin{figure}[]
\includegraphics[width=8cm,angle=0]{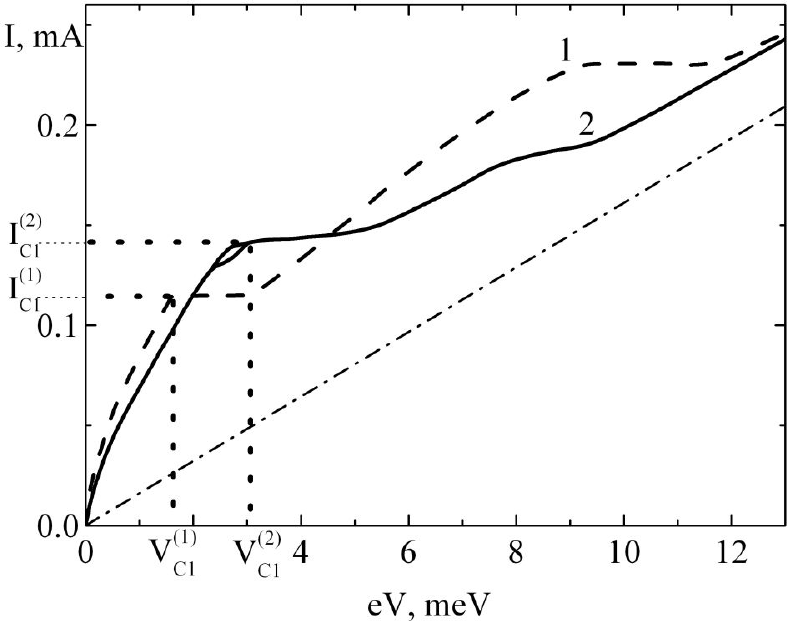}
\caption[]{Stepwise variations in the IVC of an $NbSe_2-Cu$  contact at temperatures $T=1.85$  (1) and $3.22\ K$  (2). The dot-and-dash line is parallel to IVC for large values of $V$ $\left( {{R}_{N}}=60\ \Omega  \right)$.}
\label{Fig3}
\end{figure}

Figure \ref{Fig3} shows the IVC of another junction, viz., $NbSe_2-Cu$, at different temperatures. As the temperature \textbf{decreases}, the "critical" current ${{I}_{c1}}$  and power ${{P}_{c1}}={{I}_{c1}}{{V}_{c1}}$ corresponding to a transition from the first to the second step decrease instead of increasing, as would be the case if the IVC jumps were due to the trivial destruction of superconductivity as a result of heating or the attainment of the critical current density along certain current paths. The observed temperature dependence of the IVC points towards a significant role of nonequilibrium quasiparticles. Among other things, the concentration of such particles is determined by the relaxation time which increases with decreasing temperature. Consequently, the "critical" concentration of nonequilibrium quasiparticles, which is one of the reasons behind the phase lamination in a nonequilibrium superconductor at low temperatures, is attained at a lower injection rate (lower ${{I}_{c1}}$ and ${{P}_{c1}}$).

It is interesting to note that the normal increment (${{R}_{2}}-{{R}_{1}}$) in the differential resistance at the quasilinear region corresponding to the second step is double the increment (${{R}_{1}}-{{R}_{0}}$). Since the resistance ${{R}_{n}}$  for small values of $2n{{l}_{E}}/d$  is proportional to $n$  (see \eqref{eq__1}), it can be surmised that the resulting structure will be identical to that of the phase slip centers. Indeed, for $eV\gg \Delta $, the main part of the current in the superconductor near the $N-S$  boundary is transported by the quasiparticle excitations which recombine to form pairs at a distance ${{l}_{E}}$ from the constriction. If the current density $j(\mathbf{r})$  at this distance is less than ${{j}_{c}}$, such a state is stable. Otherwise, a narrow layer (phase slip plane or PSP) is formed at a distance $2{{l}_{E}}$  from the constriction. This layer has a suppressed order parameter $\Delta $  and generates new quasiparticles capable of transporting the "supercritical" current further into the bulk of the superconductor to a depth $3{{l}_{E}}$. Such a process is repeated with increasing current until the spatial region covered by it exceeds the current spread region of the order $d$. It is obvious that the sharply steplike nature of the IVC may be observed only under the condition ${{l}_{E}}<d$, and the number of observed jumps is of the order $d/2{{l}_{E}}$. The contribution of the superconducting face to the contact resistance is given by
\begin{equation} \label{eq__1}
{{R}_{n}}=\left( \rho /\pi d \right)\operatorname{tanh}^{-1}2n{{l}_{E}}/d=2\rho n{{l}_{E}}/{{d}^{2}},\ \ \ \left( n{{l}_{E}}\ll d \right)
\end{equation}
($n$  is an odd integer).
It follows hence that the resistance of the entire contact for large values of $V$ and at the first step are given, respectively, by
\begin{equation} \label{eq__2}
{{R}_{\infty }}=\rho /2d+{{R}_{0}};\ \ {{R}_{1}}=2\rho {{l}_{E}}/{{d}^{2}}+{{R}_{0}}
\end{equation}
while the ratio $d/{{l}_{E}}$  is defined as
\begin{equation} \label{eq__3}
d/{{l}_{E}}=\frac{4}{\pi }\left( \frac{{{R}_{\infty }}-{{R}_{0}}}{{{R}_{1}}-{{R}_{0}}} \right)
\end{equation}
The contact diameter can be estimated from ${{R}_{\infty }}$ \eqref{eq__2}. Unfortunately, the exact value of the resistivity $\rho$ in the region adjoining the contact is now known. It may be much larger than ${{\rho }_{\bot }}\sim {{10}^{-3}}\ \Omega \cdot cm$ measured for a bulk sample along the $c$-axis. Of the two different $NbSe_2-Cu$ contacts whose characteristics are presented in Figs. \ref{Fig1} and \ref{Fig3}, we choose the latter for numerical estimates since it has a lower resistance, and hence a value of $\rho$  closer to the bulk value, for nearly the same number of steps on the IVC (and for the same value of the ratio $d/{{l}_{E}}$). Substituting ${{R}_{\infty }}=62\ \Omega $  and ${{R}_{0}}=10\ \Omega $ into \eqref{eq__2}, we obtain $d\gtrsim 960\ \mathrm{\AA}$, ${{l}_{E}}=145\ \mathrm{\AA}$. The coherence length in $NbS{{e}_{2}}$ is quite small: ${{\xi }_{\parallel }}=77\ \mathrm{\AA}$, ${{\xi }_{\bot }}=23\ \mathrm{\AA}$. Hence the formation of narrow PSP layers with a suppressed gap (Fig. \ref{Fig2} is quite possible.

\begin{figure}[]
\includegraphics[width=8cm,angle=0]{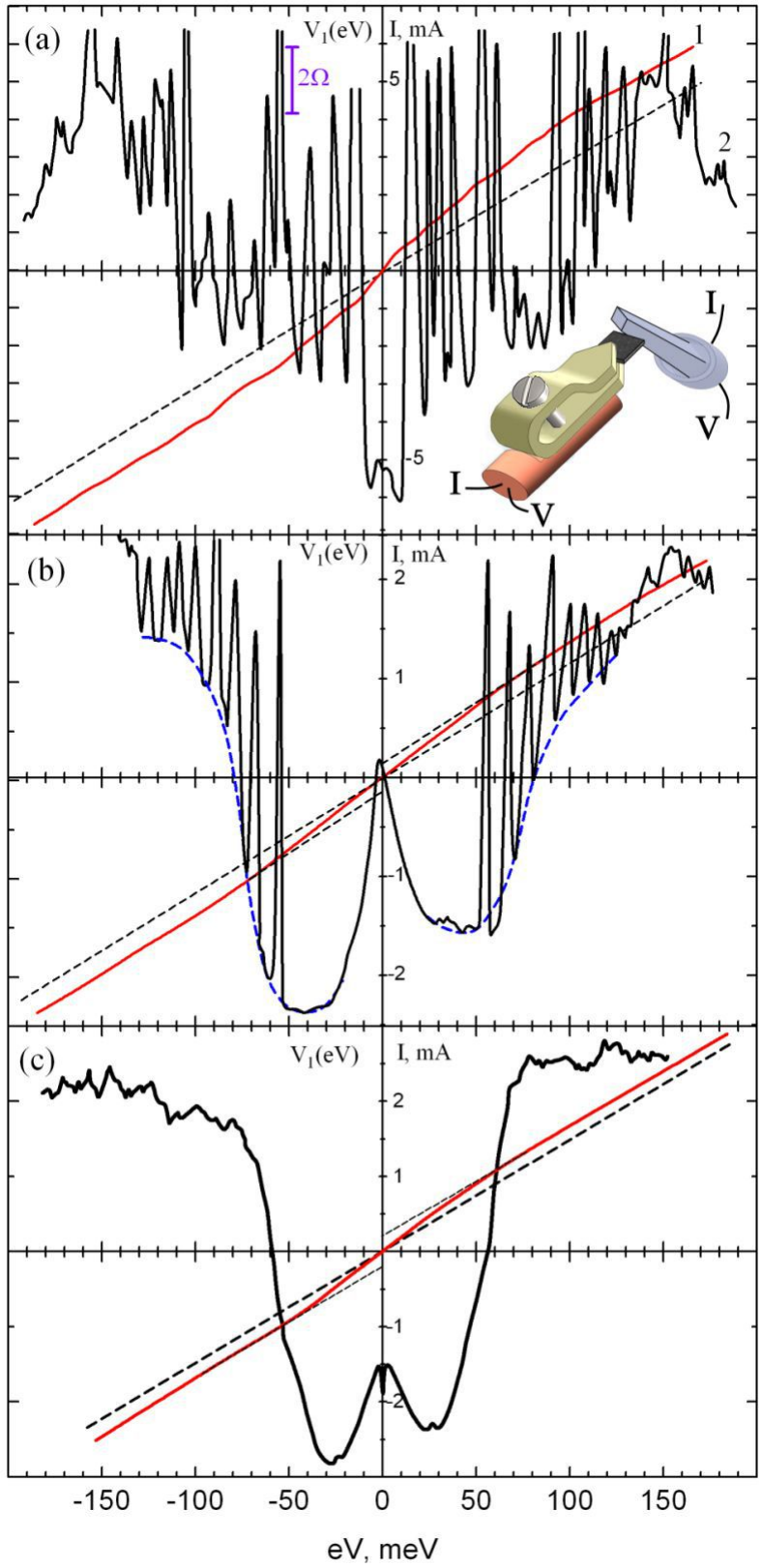}
\caption[]{(a)\ -\ stepwise structure of IVC for $YBaSrCuO-Ag$ point contacts at $T=4.2\ K$  and ${{T}_{N}}=336\ \Omega $:
1 - $I(eV)$, \\2 - ${{V}_{1}}(eV)\sim dV/dI$;\\
(b),(c)\ -\ IVC and their first derivatives ${{V}_{1}}(eV)$ $\left( {{R}_{N}}=83\ \Omega  \right)$(b)   and $68\ \Omega $(c)).}
\label{Fig4}
\end{figure}

\textbf{2.} $\textit{\textbf{YBa}}_{\textit{\textbf{1.25}}}\textit{\textbf{Sr}}_{\textit{\textbf{0.75}}}\textit{\textbf{Cu}}_{\textit{\textbf{3}}}\textit{\textbf{O}}_{7-\delta}$. Obviously, the number of observed steps must increase as the inequality $d>{{l}_{E}}>{{\xi }_{0}}$, becomes more stringent. However, in this case $d$ must not exceed the inelastic mean free path of phonons, since the thermal regime will set in otherwise. In this connection, it is interesting to use high-temperature superconductors which have record-low values of ${{\xi }_{0}}$  and $l_{\varepsilon}$. Figure \ref{Fig4}a
shows the IVC of a point contact formed by $YB{{a}_{1.25}}S{{r}_{0.75}}C{{u}_{3}}{{O}_{7-\delta }}$ and silver as a result of first touches between the electrodes in liquid helium. The contact geometry is shown in the inset. It can be seen that the IVC comprises a large number of (about twenty) steps with a gradually increasing slope $dV/dI$. For subsequent touches between the electrodes, the stepwise structure of the IVC is indistinguishable but its existence can be clearly seen in the first derivative plots (Fig. \ref{Fig4}b). The number of steps decreases in comparison with the preceding contact and their arrangement reveals a regularity manifested in a smooth decrease in the distance between spikes in ${{V}_{1}}(eV)\sim dV/dI$ with increasing step number. The envelope ${{V}_{1}}(eV)$, shown by a dashed line, has minima at $\pm 42\ meV$  which can be interpreted as manifestations of the energy gap $\Delta$ due to a small "tunneling" component of the current emerging as a result of the reflection of a part of the electrons at the physical boundary between the superconductor and the normal metal.
The nonlinearity of IVC for $V=0$  vanishes at $T=70\ K$  (this temperature corresponds to the end of the resistive transition of the bulk sample into the superconducting state). Taking this temperature as $T_c$, we obtain $2\Delta /k{{T}_{c}}\simeq 14$, which is anomalously large (as compared to the value given by the BCS theory) and is nearly equal to the frequently encountered values in the tunnel- or point-contact studies of high-temperature superconductors.

In the same series of experiments, subsequent mutual displacements of electrodes lead to the disappearance of the above-mentioned structure of $dV/dI$  caused, in our opinion, by the discrete nature of the displacement of the ${S}'S$  boundary, and also to a considerable decrease in the gap value $\Delta $  to about $25\ meV$, which points towards a partial degradation of superconducting properties near the contact (Fig. \ref{Fig4}c).

Thus, it can be concluded that the regular stepwise nature of IVC of point contacts involving superconductors with short ${{\xi }_{0}}$  and ${{l}_{E}}$ is observed only for an unperturbed superconductor with a perfect lattice. The disorder, which is created during the formation of the point contact and is manifested, among other things, in a reduced (or averaged over different directions) gap, prevents the formation of a regular system of phase boundaries $S'S$  in the contact region. This explains the absence of a regular structure of $dV/dI(V)$  in the studies involving HTS ceramic materials.

\begin{figure}[]
\includegraphics[width=8cm,angle=0]{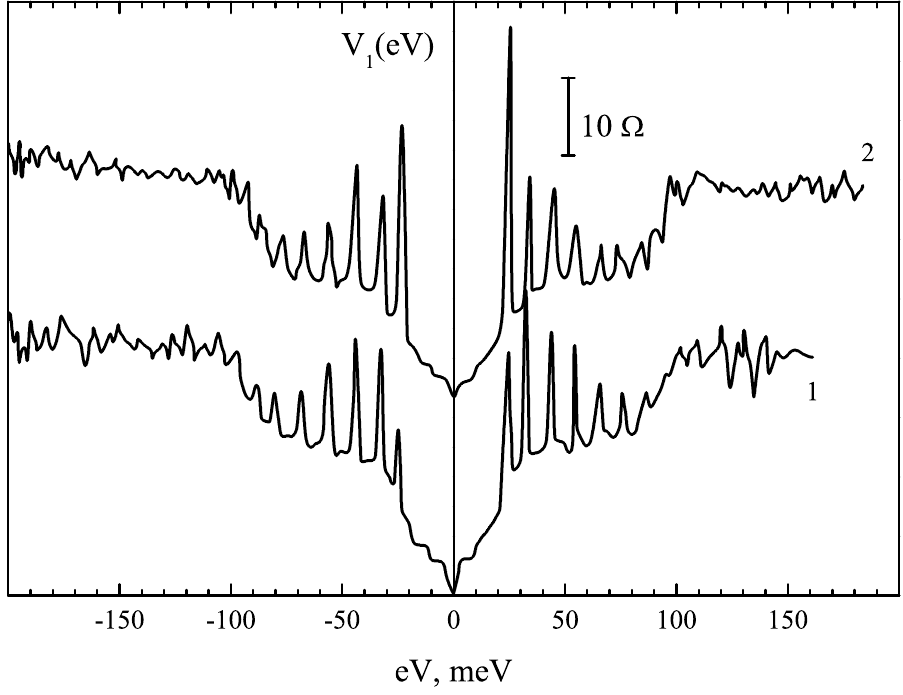}
\caption[]{Periodic structure of the $dV/dI$ characteristic of a $YBaSrCuO-Ag$ contact (with $\left( {R_N}\simeq105\ \Omega  \right)$ at $T=10$ (1) and $20\ K$ (2).}
\label{Fig5}
\end{figure}

\begin{figure}[]
\includegraphics[width=8cm,angle=0]{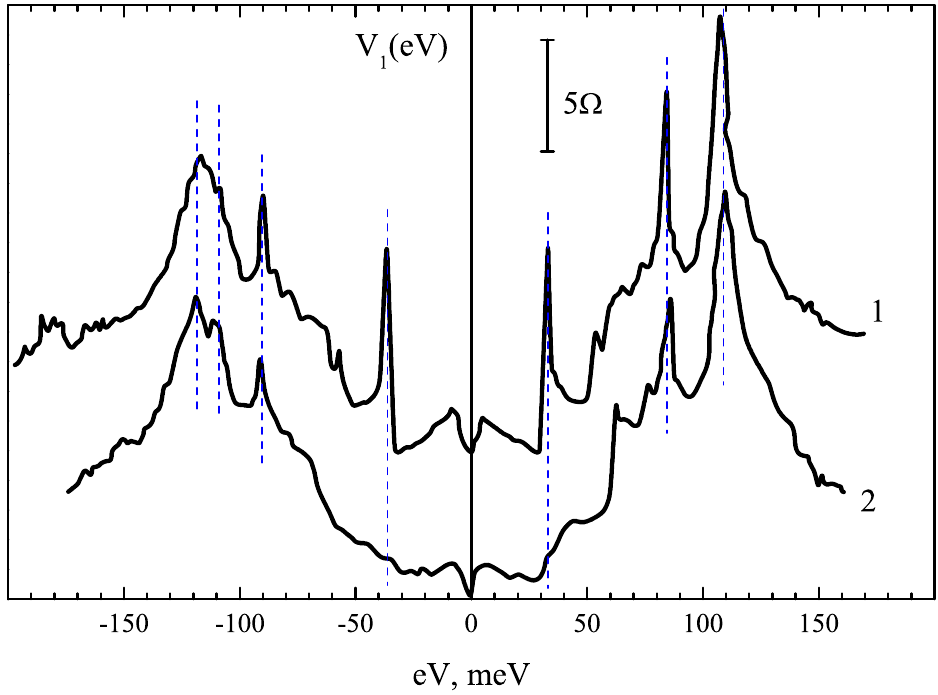}
\caption[]{Effect of magnetic field on the peak-like peculiarities of the IVC derivative of a $YBaSrCuO-Ag$ contact at $T=4.2\ K$, $R_N=30\ \Omega$: $H=0$  (1) and $H=40\ kOe$ (2).}
\label{Fig6}
\end{figure}

Figure \ref{Fig5}
shows the ${{V}_{1}}\left( eV \right)$ structure for another contact formed by the same single crystal, recorded at two different temperatures. As the temperature increases, a small displacement of some peaks towards higher voltages can be observed. This displacement is analogous to the one discussed above for $NbSe_2$ (Fig. \ref{Fig3}). Such an effect is observed more clearly upon the application of an external magnetic field (Fig. \ref{Fig6}).
The characteristics shown in Fig. \ref{Fig6} correspond to a polycrystalline sample obtained by cooling the melt. This may be the reason behind the irregular structure of ${{V}_{1}}\left( eV \right)$. It can be seen that the magnetic field considerably decreases the intensity of peaks. The ${{V}_{1}}\left( eV \right)$ peaks are broadened and displaced towards \textbf{higher} energies.

Hence, it can be concluded on the basis of the temperature and field dependences presented above that the sharp peaks in the ${{V}_{1}}\left( eV \right)$ dependence are not associated with the usual mechanism of the violation of current paths in inhomogeneous HTS samples due to heating or a local increase in the critical current density. Using formulas \eqref{eq__1}-\eqref{eq__3}, we can estimate the order of magnitude of the effective diffusion depth ${{l}_{E}}$ quasiparticle excitations in an HTS. Using the value ${{\rho }_{\parallel }}=200\cdot {{10}^{-6}}\ \Omega \cdot cm$ for ${{R}_{\infty }}=100\ \Omega $, we obtain the current diameter $d\approx {{10}^{-6}}\ cm$. Proceeding from Fig. \ref{Fig4}b as a typical case, the number of layers of thickness ${{l}_{E}}$ in the current concentration region is estimated at about 10. Consequently, ${{l}_{E}}\approx d/10\sim {{10}^{-7}}cm$, which is smaller than the corresponding value for $NbSe_2$ by an order of magnitude.

Let us clarify the meaning of the effective diffusion depth ${{l}_{E}}$ of quasiparticles. Quasiparticles with an excess energy relative to the electrochemical pair potential (right up to $eV$) are injected, into the HTS layer adjoining the $N-S$ boundary. However, if $eV\gg \Delta $, the quasiparticles rapidly relax to states above the energy gap, and hence the value of  $l_E$ is mainly determined by the quasiparticles having an excitation energy of the order of $\Delta $. For such quasiparticles, it is reasonable to use the values $\rho (T)$ at $T=T_c$, as was indeed done by us in this paper.

For the sake of comparison, it is worthwhile to consider the results of point-contact investigations in $Ta$, \cite{Yanson3} whose EPI function is well known. As a rule, the $S-c-N$ contacts with noble metals reveal only one nonequilibrium singularity. It was mentioned above the this singularity has a rather weak intensity and can therefore be associated with the formation of an ${S}'S$ boundary at the periphery of the current spread region. In many cases, the nucleation of this singularity was ascribed to the characteristic phonon energies on the $eV$ axis.
It is also worth noting that an analysis of the point contacts formed by the $Nb_3Sn$ single crystals and $Cu$ also reveals a stepwise structure of the IVC whose main features are analogous to those for the $NbSe_2$ contacts described above.

\end{document}